\documentclass[twocolumn,aps,english,prl,nofootinbib]{revtex4}
\usepackage{mathrsfs}
\usepackage{graphicx}
\usepackage{amsmath, amssymb}
\usepackage{babel}
\usepackage{color}
\usepackage{slashed}

\newcommand{\ali}{&\!\!}

\begin{document}

\title{Novel Soft-Pion Theorem for Long-Range Nuclear Parity 
Violation}

\author{Xu Feng$^{a,b,c}$}
\author{Feng-Kun Guo$^{d,e}$}
\author{Chien-Yeah Seng$^{f}$}

\affiliation{$^{a}$School of Physics and State Key Laboratory of Nuclear Physics and 
Technology,\\ Peking University, Beijing 100871, China}
\affiliation{$^{b}$Collaborative Innovation Center of Quantum Matter, Beijing 100871, China}
\affiliation{$^{c}$Center for High Energy Physics, Peking University, Beijing 
100871, China}
\affiliation{$^{d}$CAS Key Laboratory of Theoretical Physics,
            Institute of Theoretical Physics,\\ Chinese Academy of Sciences,
            Beijing 100190, China}
\affiliation{$^{e}$School of Physical Sciences,
            University of Chinese Academy of Sciences,
            Beijing 100049, China}
\affiliation{$^{f}$INPAC, Shanghai Key Laboratory for Particle Physics and Cosmology, \\
	MOE Key Laboratory
	for Particle Physics, Astrophysics and Cosmology,  \\
	School of Physics and Astronomy, Shanghai Jiao-Tong University, Shanghai 200240, China}

\date{\today}

\begin{abstract}

The parity-odd effect in the Standard Model weak neutral current reveals itself 
in the long-range parity-violating nuclear potential generated by the pion 
exchanges in the $\Delta I=1$ channel with the parity-odd pion-nucleon coupling 
constant $h_\pi^1$. Despite decades of experimental and 
theoretical efforts, the size of this coupling constant is still not 
well-understood. In this Letter we derive a soft-pion theorem relating
$h_\pi^1$ and the neutron-proton mass-splitting induced by an 
artificial parity-even counterpart of the $\Delta I=1$ weak Lagrangian, and 
demonstrate that {\color{black}the theorem still holds exact at the next-to-leading order} in chiral perturbation 
theory. A considerable amount of simplification is 
expected in the study of $h_\pi^1$ by using either lattice or other QCD models 
following its reduction from a parity-odd proton-neutron-pion matrix element to 
a simpler spectroscopic quantity. {\color{black} The theorem paves the way to much more precise calculations of $h_\pi^1$, and thus a quantitative test of the strangeness-conserving neutral current interaction of the Standard Model is foreseen. }

\end{abstract}

\maketitle

The study of the Standard Model~(SM) Parity~(P)-violation in non-leptonic 
processes is extremely difficult due to the overwhelming background of strong 
interaction governed by Quantum Chromodynamics~(QCD). Nevertheless, it is 
essential {\color{black}in order to better understand} the general properties of the hadronic 
weak interaction and {\color{black}shed light on many unresolved puzzles such as the unexpectedly large violation of Hara's theorem in the
hyperon weak radiative decays and the failure to simultaneously fit the S- 
and P-wave amplitudes of the hyperon decays.}
%zAlthough lattice QCD is 
%able to account for some of these anomalies~\cite{Boyle:2012ys,Blum:2015ywa}, a 
%clear physical picture to the reason of their occurrence remains absent; in 
%particular, we do not know if they are associated to the special role of the 
%strange quark in QCD or if there is a deeper reason lying within the interplay 
%of the strong and weak interaction. The study of strangeness-conserving weak 
%hadronic process may therefore possibly shed light to this 
%question~\cite{RamseyMusolf:2006dz}.
{\color{black}The P-violation in the strangeness-conserving, $\Delta I=1$ nucleon-nucleon 
interaction is a perfect ground to study the properties of neutral weak current in hadronic systems which is otherwise poorly-constrained as it cannot be probed in the usual strangeness-changing weak processes due to the absence of a tree-level flavor-changing neutral current \cite{Gardner:2017xyl}. Ongoing P-violation experiments with an unprecedented level of precision (see discussion below) call for a new round of theoretical study of the hadronic P-violation so that the experimental results can be utilized to their largest extent in testing our current understanding of the SM hadronic weak interaction.}

On the 
theory side, the paper by Desplanques, Donoghue and 
Holstein~(DDH)~\cite{Desplanques:1979hn} formulated both the P-conserving (due 
to the strong interaction) and the P-violating (due to the weak interaction including 
all $\Delta I=0,1,2$ channels) nucleon-nucleon interaction in terms of the 
single exchange of the lowest-lying light mesons, i.e. $\pi,\rho$ and $\omega$. 
This description forms the basis of many experimental analyses. More recently, 
the description of the P-violating nucleon-nucleon forces and the associated 
currents 
has been based on effective field theory~(EFT) frameworks such as the pionless 
EFT~\cite{Griesshammer:2010nd, Schindler:2013yua} or the chiral 
EFT~\cite{Zhu:2004vw, Viviani:2014zha, deVries:2015gea}. Experimental progress 
has been made as well, although mainly in the last decades of the previous 
century, as exemplified by measurements of the P-violating longitudinal 
analyzing 
powers~(LAPs)~\cite{Eversheim:1991tg,Kistryn:1987tq,Nagle:1978vn,Berdoz:2001nu, 
Lang:1985jv}, the gamma-ray asymmetries 
\cite{Cavaignac:1977uk,Gericke:2011zz,Adelberger:1983zz,Elsener:1987sx, 
Elsener:1984vp}, and the gamma-ray circular 
polarization~\cite{Lobashov:1972fwg,Knyazkov:1984lzj,Barnes:1978sq,Bini:1985zz, 
Ahrens:1982vfn, Page:1987ak}. {For recent reviews, we refer to 
Refs.~\cite{RamseyMusolf:2006dz,deVries:2015gea,Gardner:2017xyl}.}

A long-standing problem in the field of hadronic P-violation is the theoretical  
determination of the P-odd hadronic coupling constants. Since the underlying 
weak operators and their Wilson coefficients are rather well-known, the 
outstanding problem is to compute the associated hadronic matrix elements 
directly with non-perturbative methods or to fit them to data. In particular, 
the P-violating, $\Delta I=1$ pion-nucleon coupling $h_\pi^1$ has attracted much attention as 
it is formally the single leading-order (LO) operator in the chiral EFT 
framework~\cite{Kaplan1993}. As such, it is expected to dominate the long-range 
part of the P-violating nucleon-nucleon potential and the resulting P-odd 
phenomenology in various processes. The above conclusion, however, depends 
crucially on the actual size of $h_\pi^1$, and if it turns out smaller than 
originally expected other P-violating hadronic interactions can become dominant. 
Such interactions are described by the P-odd couplings between nucleons and 
heavier mesons (like the $\rho$ and $\omega$) in the DDH approach or as 
the P-odd nucleon-nucleon contact~\cite{Zhu:2004vw, Girlanda:2008ts} and 
derivative pion-nucleon~\cite{deVries:2014vqa} interactions in the EFT 
language. 

The simplest order-of-magnitude estimate of the size of $h_\pi^1$ is based on 
the naive dimensional analysis which gives $h_\pi^1  \sim \mathcal O(G_F 
F_\pi \Lambda_\chi) \sim 10^{-6}$ in terms of the Fermi constant $G_F$, the pion 
decay constant $F_\pi$, and the chiral-symmetry-breaking scale $\Lambda_\chi$. 
This estimate does not capture the potential suppression due to powers of 
{$\sin\theta_W$} or large $N_c$ 
arguments~\cite{Phillips:2014kna,Gardner:2017xyl}. 
In the DDH paper, a ``reasonable range" for $h_\pi^1$ is given as $(0-11)\times 
10^{-7}$ together with a ``best value" around $4.6\times 10^{-7}$ based on a 
quark model and SU(6) flavor-spin symmetry. Other phenomenological studies of 
$h_\pi^1$ include the use of quark 
models~\cite{Dubovik:1986pj,Feldman:1991tj,Hyun:2016ddn}, Skyrme 
models~\cite{Kaiser:1988bt,Kaiser:1989fd,Meissner:1998pu}, and QCD sum 
rules~\cite{Henley:1995ad,Lobov:2002xb}. In general, both the quark model and 
QCD sum rules predict an order of $10^{-7}$ for $h_\pi^1$; meanwhile, early chiral 
	Skyrmion approaches~\cite{Kaiser:1988bt,Kaiser:1989fd} predict
$\mathcal{O}(10^{-8})$ but subsequent work~\cite{Meissner:1998pu} gives 
$10^{-7}$, all rather small values that 
are difficult to probe experimentally.

On one hand, such small values are in good agreement with the absence of a 
P-odd signal in the $\gamma$-emission from ${}^{18}$F which gives an upper 
bound $h_\pi^1 \leq 1.3\times
10^{-7}$~\cite{Adelberger:1983zz,Haxton:1981sf,Page:1987ak}. On the other hand, 
the size of the Cs anapole moment~\cite{Wood:1997zq,Haxton:2001mi} indicates a 
larger $h_\pi^1$, and the same holds 
for the measurement of the LAP in proton-alpha scattering 
\cite{Lang:1985jv,Roser:1985rs}. The interpretation of the latter experiments, 
however, suffers from significant theoretical uncertainties due to the 
complicated systems involved in the experiments. A more promising approach seems 
to rely only on experiments involving a few nucleons where the nuclear theory is 
under better control. In this light, 
Refs.~\cite{deVries:2013fxa,Viviani:2014zha} tried to extract $h_\pi^1$ from 
measurements of the proton-proton LAP. Unfortunately, $h_\pi^1$ contributes only 
to the proton-proton scattering at the subleading order, leading to a large 
uncertainty 
on the extraction $h^1_\pi = (1.1\pm 2.0 )\times 10^{-6}$. More promising is 
the 
extraction of $h_\pi^1$ from the upcoming measurement of the gamma-ray asymmetry 
in the neutron capture on the proton by the NPDGamma 
Collaboration~\cite{Fry:2016esm, deVries:2015pza}. 

{\color{black}In view of these experimental efforts to extract $h_\pi^1$, there is the need to calculate its value reliably using, e.g., lattice QCD, in order to quantitatively test the strangeness-conserving neutral current aspect of the SM.}
So far the only direct lattice QCD calculation of 
$h_\pi^1$ was attempted in Ref.~\cite{Wasem:2011zz} by studying a three-point 
correlation function. The result was incomplete partially due to its inability 
to extract signals from the so-called quark-loop diagram, which suffered from a 
too small signal-to-noise ratio. The existence of an explicit pion in the final 
state also brought about extra technical complications. For example an extra 
total-derivative operator with an unknown coefficient must be introduced for 
the insertion of energy into the weak vertex. 

In this Letter we propose a new starting point for the theoretical 
investigation 
of $h_\pi^1$ by deriving a soft-pion theorem. This theorem relates $h_\pi^1$ to 
the neutron-proton mass splitting induced 
by an artificial P-even counterpart of the $\Delta I=1$ weak Lagrangian. 
This approach is parallel to one of the techniques in the 
study of P-and-time-reversal-odd pion-nucleon coupling $\bar{g}_\pi^i$, 
which also attempts to relate $\bar{g}_\pi^i$ to the nucleon mass shifts 
generated by the underlying P-even operators. Such relations were first derived 
using the current algebra~\cite{Crewther:1979pi,Pospelov:2005pr} and later 
refined 
under the framework of chiral perturbation 
theory~(ChPT)~\cite{deVries:2015una,deVries:2012ab,Mereghetti:2015rra,
Seng:2016pfd,
deVries:2016jox,Cirigliano:2016yhc}. In its application to $h_\pi^1$, we find 
that the simple matching relation derived in LO ChPT is 
{\color{black} exactly preserved by all corrections at the next-to-leading-order (NLO), i.e. $\mathcal{O}(M_\pi^2/\Lambda_\chi^2)$ with $M_\pi$ the pion mass and $\Lambda_\chi\approx1$~GeV the chiral symmetry breaking scale,}
including both the one-loop and low-energy constant (LEC) contributions, which 
is a unique feature not shared by $\bar{g}_\pi^i$. Hence, the accuracy of such a
simple matching relation is expected to be better than one percent. Considerable 
advantages are expected by studying the neutron-proton mass splitting instead 
of $h_\pi^1$ itself using either lattice QCD or other nonperturbative 
approaches.

We start by reviewing the underlying physics of the flavor-conserving nuclear P-violation in the SM. Well below the electroweak scale, the 
$W$ or $Z$-bosons can be integrated out in exchange of four-quark operators 
in the form of current-current products. In the limit of vanishing Cabibbo 
angle $\theta_C$, the charged current 
will not contribute to the $\Delta I=1$ parity violation and in reality it is 
suppressed by $\sin^2\theta_C\simeq0.05$. Hence, {\color{black}the flavor-conserving $\Delta 
I=1$ parity-violating nuclear processes serve as a unique probe to test the otherwise poorly-constrained neutral current interaction of the SM}. If one considers only the three lightest quarks, the effects of P-violation in 
the $\Delta S=0$, $\Delta I=1$ channel are carried by seven independent P-odd 
four-quark operators which take the following 
form~\cite{Tiburzi:2012hx}\footnote{There is another operator $\theta_4$ 
defined in Ref.~\cite{Kaplan:1992vj}, but it is not independent from 
$\{\theta_1,\theta_2,\theta_3\}$.}:
\begin{equation}
\mathcal{L}_{\mathrm{PV}}^{w}=-\frac{G_{F}}{\sqrt{2}}\frac{\sin^{2}\theta_{W}}{3
}\sum_{i}\left(C_{i}^{(1)}\theta_{i}+S_{i}^{(1)}\theta_{i}^{(s)}\right),
\label{eq:Lweak}
\end{equation} 
where
\begin{eqnarray}
\theta_{1} \ali = \ali 
\bar{q}_a\gamma^{\mu}q_a\bar{q}_b\gamma_{\mu}\gamma_{5}\tau_{3}q_b\, ,~~
\theta_{2} =
\bar{q}_{a}\gamma^{\mu}q_{b}\bar{q}_{b}\gamma_{\mu}\gamma_{5}\tau_{3}q_{a}\, ,
\nonumber \\
\theta_{3} \ali = \ali 
\bar{q}_a\gamma^{\mu}\gamma_{5}q_a\bar{q}_b\gamma_{\mu}\tau_{3}q_b\, ,\nonumber 
\\
\theta_{1}^{(s)} \ali = \ali 
\bar{s}_a\gamma^{\mu}s_a\bar{q}_b\gamma_{\mu}\gamma_{5}\tau_{3}q_b\, ,~~
\theta_{2}^{(s)} =
\bar{s}_{a}\gamma^{\mu}s_{b}\bar{q}_{b}\gamma_{\mu}\gamma_{5}\tau_{3}q_{a}\, ,\nonumber\\
\theta_{3}^{(s)} \ali=\ali
\bar{s}_a\gamma^{\mu}\gamma_{5}s_a\bar{q}_b\gamma_{\mu}\tau_{3}q_b\, ,~~
\theta_{4}^{(s)} =
\bar{s}_{a}\gamma^{\mu}\gamma_{5}s_{b}\bar{q}_{b}\gamma_{\mu}\tau_{3}q_{a}\,.
\end{eqnarray}
Here $q=(u,d)^T$ is the quark isospin doublet field, $s$ is the strange 
quark field, $\{a,b\}$ are color indices and 
$\theta_W$ is the weak mixing angle. The mixing of these operators under 
one-loop perturbative QCD corrections introduces a scale dependence to the 
Wilson coefficients 
$\{C_i^{(1)},S_i^{(1)}\}$~\cite{Dai:1991bx,Kaplan:1992vj,Tiburzi:2012hx}.
{\color{black}The uncertainties of the Wilson coefficients are relatively under control: for instance, the higher-order corrections to the LO QCD running are generally of the order of 10-20\%~\cite{Tiburzi:2012hx}.}
 % Here 
% we quote their 
% values at the scale $\Lambda_\chi\approx1$\,GeV as 
% calculated in Ref. \cite{Tiburzi:2012hx}:
% \begin{eqnarray}
% C^{(1)}(\Lambda_{\chi}) \ali = \ali \left(\begin{array}{ccc}
% -0.055 & 0.810 & -0.627 \end{array}\right),\nonumber \\
% S^{(1)}(\Lambda_{\chi}) \ali = \ali \left(\begin{array}{cccc}
% 5.09 & -2.55 & 4.51 & -3.36\end{array}\right).\label{eq:Wilson}
% \end{eqnarray}

The $\Delta I=1$ P-violating coupling constants in either the DDH formalism or the 
EFT description are then just the QCD matrix elements of the Lagrangian 
\eqref{eq:Lweak} at the scale $\mu\approx \Lambda_\chi$ with respect to  
appropriate external hadronic states. The aim of this Letter is to relate these 
matrix elements to another set of P-even matrix elements with fewer external 
states. {\color{black}The easiest way to understand this formalism is to realize that the partially-conserved axial current (PCAC) relation relates matrix elements with and without a soft external pion:
\begin{equation}
\lim_{p_\pi\rightarrow 0}\left\langle 
N'\pi^i\right|\mathcal{L}_\mathrm{PV}^w\left|N\right\rangle\approx\frac{i}{F_\pi}\left\langle 
N'\right|\left[\mathcal{L}_\mathrm{PV}^w,\hat{Q}^i_A\right]\left|N\right\rangle , \label{eq:PCAC}
\end{equation}	
where $\hat{Q}_A^i$ is the axial charge; the right hand side can be further reduced to a flavor-diagonal matrix element through the Wigner--Eckart theorem. This observation inspires us to construct a P-even chiral partner of $\mathcal{L}_\mathrm{PV}^w$ as follows}:
\begin{equation}
\mathcal{L}_{\mathrm{PC}}^{w}=-\frac{G_{F}}{\sqrt{2}}\frac{\sin^{2}\theta_{W}}{3
}\sum_{i}\left(C_{i}^{(1)}\theta_{i}'+S_{i}^{(1)}\theta_{i}^{(s)\prime}
\right)\label{eq:Lageven}
\end{equation}
where 
\begin{eqnarray}
\theta_{1}' \ali = \ali 
\bar{q}_a\gamma^{\mu}q_a\bar{q}_b\gamma_{\mu}\tau_{3}q_b\,, ~~
\theta_{2}' =
\bar{q}_{a}\gamma^{\mu}q_{b}\bar{q}_{b}\gamma_{\mu}\tau_{3}q_{a}\,,  \\
\theta_{3}' \ali = \ali 
\bar{q}_a\gamma^{\mu}\gamma_{5}q_a\bar{q}_b\gamma_{\mu}\gamma_{5}\tau_{3}
q_b \,,\nonumber \\
\theta_{1}^{(s)\prime} \ali = \ali 
\bar{s}_a\gamma^{\mu}s_a\bar{q}_b\gamma_{\mu}\tau_{3}q_b\,,~~
\theta_{2}^{(s)\prime} = 
\bar{s}_{a}\gamma^{\mu}s_{b}\bar{q}_{b}\gamma_{\mu}\tau_{3}q_{a}\,,\nonumber \\
\theta_{3}^{(s)\prime} \ali = \ali 
\bar{s}_a\gamma^{\mu}\gamma_{5}s_a\bar{q}_b\gamma_{\mu}\gamma_{5}\tau_{3}
q_b \,,~~
\theta_{4}^{(s)\prime} =
\bar{s}_{a}\gamma^{\mu}\gamma_{5}s_{b}\bar{q}_{b}\gamma_{\mu}\gamma_{5}\tau_{3}
q_{a}\,.\nonumber
\label{eq:LPeven}
\end{eqnarray} 

% Notice that the Wilson coefficient of each P-even operator is chosen 
% to be identical to its P-odd counterpart on purpose. 

Regardless of the DDH formalism or the (pionful) EFT description, the 
long-range P-violating nuclear potential always features pion-exchanges. If we 
write $N=(p,n)^T$ as the nucleon isospin doublet, then there are four kinds of 
$NN\pi$ couplings one could write down in terms of the isospin decomposition: 
$\bar{N}\vec{\tau}\cdot\vec{\pi}N$, $\pi^0\bar{N}N$, 
$\bar{N}(\vec{\tau}\times\vec{\pi})^3N$ and 
$\bar{N}(\vec{\tau}\cdot\vec{\pi}-3\pi^0\tau_3)N$ where the first term has 
$\Delta I=0$, the second and third terms have $\Delta I=1$ and the last term 
has $\Delta I=2$. Since Barton's theorem~\cite{Barton:1961eg} excludes the 
possibility of exchanging neutral pseudoscalars in the CP-conserving limit, 
the only available structure is $\bar{N}(\vec{\tau}\times\vec{\pi})^3N$ which 
has $\Delta I=1$ and is therefore dominated by the neutral current contribution. 
The LO P-odd pion-nucleon coupling term can thus be written as
\begin{equation}
\mathcal{L}_{\mathrm{PV}}^{w}=-\frac{h_{\pi}^{1}}{\sqrt{2}}\bar{N}(\vec{\tau}
\times\vec{\pi})^{3}N + \ldots
 = i\,h_\pi^1 \left( \bar n p \pi^- - \bar p n \pi^+ \right) +\ldots 
,\label{eq:hpidef}
\end{equation}
where the pion-nucleon coupling constant $h_{\pi}^{1}$ may be expressed in terms 
of the hadronic matrix element
% \footnote{Here we choose the the Lorentz-invariant 
% normalization condition for a single-particle state: $\left\langle 
% \vec{p}'\right|\left.\!\vec{p}\right\rangle=(2\pi)^32E_p\delta^3(\vec{p}-\vec{p}
% ')$.
% % which is standard in field theory but may be different from usual lattice 
% % convention.
% },
\begin{equation}
h_{\pi}^{1}=-\frac{i}{2m_N}\lim_{p_\pi\rightarrow 0}\left\langle 
n\pi^+\right|\mathcal{L}_{\mathrm{PV}}^{w}(0)\left|p\right\rangle.
\label{eq:hpitoME}
\end{equation}
Here $m_N$ is the averaged nucleon mass. 

We shall now derive the promised soft-pion theorem using ChPT. As 
far as this work is concerned, it is 
sufficient to restrict ourselves to the SU(2) version of ChPT since 
strangeness is conserved in the weak Lagrangian of our interest. We should 
stress that this does not mean that we are disregarding the effects of the 
operators with strange fields, i.e. $\{\theta_i^{(s)}\}$; they are just having 
the same isospin structure as the non-strange operators $\{\theta_i\}$ and 
can be described by the same spurion involving only SU(2)  indices, as we 
shall demonstrate later. The LO chiral Lagrangians for QCD in the pion and nucleon sector are 
given by
\begin{eqnarray}
\mathcal{L}_\pi\ali=\ali \frac{F_0^2}{4}\mathrm{Tr}\left[\partial_\mu 
U\partial^\mu U^\dagger\right]+\frac{F_0^2B_0}{2}\mathrm{Tr}\left[
M_qU^\dagger+UM_q^\dagger\right], \nonumber\\
\mathcal{L}_N\ali=\ali \bar{N}iv\cdot\mathcal{D}N+g_A\bar{N}u_\mu S^\mu N\,,
\end{eqnarray}
respectively. {\color{black}Here we adopt the standard notations of ChPT as in Ref.~\cite{Scherer:2012xha}: $U=\exp\left\{{i\vec{\pi}\cdot\vec{\tau}/F_0}\right\}$, $u=\sqrt{U}$, and $u_\mu=i(u^\dagger\partial_\mu u-u\partial_\mu u^\dagger)$ while} $M_q=\mathrm{diag}(m_u,m_d)$ is the quark mass matrix that gives 
rise to the LO pion mass $M_\pi^2=B_0(m_u+m_d)$. In the nucleon 
sector, we adopt the heavy baryon chiral perturbation theory 
formalism~\cite{Jenkins:1990jv,Bernard:1995dp} so that the nucleon field $N$ 
appears as a massless excitation {\color{black}with four-velocity $v$ and the chiral covariant derivative $\mathcal{D}_\mu=\partial_\mu+(u^\dagger\partial_\mu u+u\partial_\mu u^\dagger)/2$}. The finite quark mass effect could be implemented to the 
baryon Lagrangian at higher orders through the matrices 
$\chi_\pm=2B_0(u^\dagger M_q u^\dagger\pm u M_q^\dagger u)$. Next we turn to the discussion of the weak chiral Lagrangian. The effects of both $\mathcal{L}_{\mathrm{PV}}^w$ and $\mathcal{L}_{\mathrm{PC}}^w$ can be
implemented into the chiral Lagrangian by means of the spurion method. To 
understand the procedure, we first combine the two Lagrangians to obtain
\begin{equation}
\mathcal{L}_{\mathrm{tot}}^{w} = 
\mathcal{L}_{\mathrm{PV}}^{w}+\mathcal{L}_{\mathrm{PC}}^{w}=-\frac{G_{F}}{\sqrt{
2}}\frac{\sin^{2}\theta_{W}}{3}\sum_{i}\left(C_{i}^{(1)}\tilde{\theta}_{i}+S_{i}
^{(1)}\tilde{\theta}_{i}^{(s)}\right),
\end{equation}
where $\tilde{\theta}_i=\theta_i+\theta_i'$ and 
$\tilde{\theta}_i^{(s)}=\theta_i^{(s)}+\theta_i^{(s)\prime}$. 
One immediately observes that the operators $\{\tilde{\theta}_i,\tilde{\theta}_i^{(s)}\}$ break the SU(2) chiral 
symmetry via the matrix $\tau_3$. Therefore, the effect of $\mathcal{L}_{\mathrm{tot}}^{w}$ can 
be implemented to the chiral Lagrangian through a Hermitian, traceless spurion,
% \begin{equation}
$X_R=u^\dagger \tau_3 u$.
% \end{equation}
The LO weak 
Lagrangian in the nucleon sector that incorporates such a spurion is simply 
\cite{Kaplan:1992vj,Beane:2002ca}
\begin{align}
\mathcal{L}_{\mathrm{tot,LO}}^{w} & =\alpha\bar{N}X_{R}N\nonumber \\
& 
=\alpha\bar{N}\tau^{3}N-\frac{\sqrt{2}i}{F_{0}}\alpha(\bar{n}p\pi^{-}-\bar{p}
n\pi^{+})+\ldots \,, \label{eq:LwLO}
\end{align}
where $\alpha$ is an unknown LEC. In the second line we have expanded the 
Lagrangian to the first power of the pion field; the first term corresponds to 
the 
neutron-proton mass splitting while the second corresponds to the P-odd 
pion-nucleon coupling. The fact that they share the same LEC $\alpha$ implies a 
relation between these two quantities:
\begin{equation}
F_\pi h_\pi^1\approx-\frac{(\delta m_N)_{4q}}{\sqrt{2}}\,,\label{eq:central}
\end{equation}
where $(\delta m_N)_{4q}$ is the neutron-proton mass splitting induced by $\mathcal{L}_{\mathrm{PC}}^w$: 
\begin{equation}
(\delta m_N)_{4q}=\frac{1}{m_N}\left\langle 
p\right|\mathcal{L}_{\mathrm{PC}}^{w}(0)\left|p\right\rangle=-\frac{1}{m_N}
\left\langle 
n\right|\mathcal{L}_{\mathrm{PC}}^{w}(0)\left|n\right\rangle.\label{eq:FH_thm}
\end{equation}

\begin{figure}
	\begin{centering}
     \includegraphics[width=0.32\linewidth]{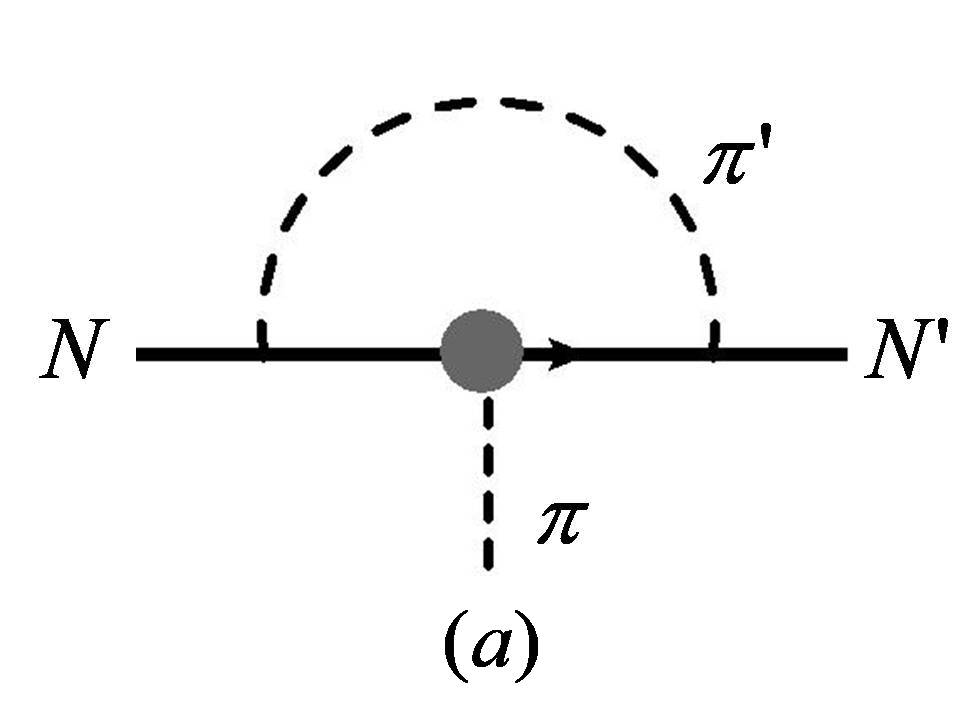}\hfill
     \includegraphics[width=0.32\linewidth]{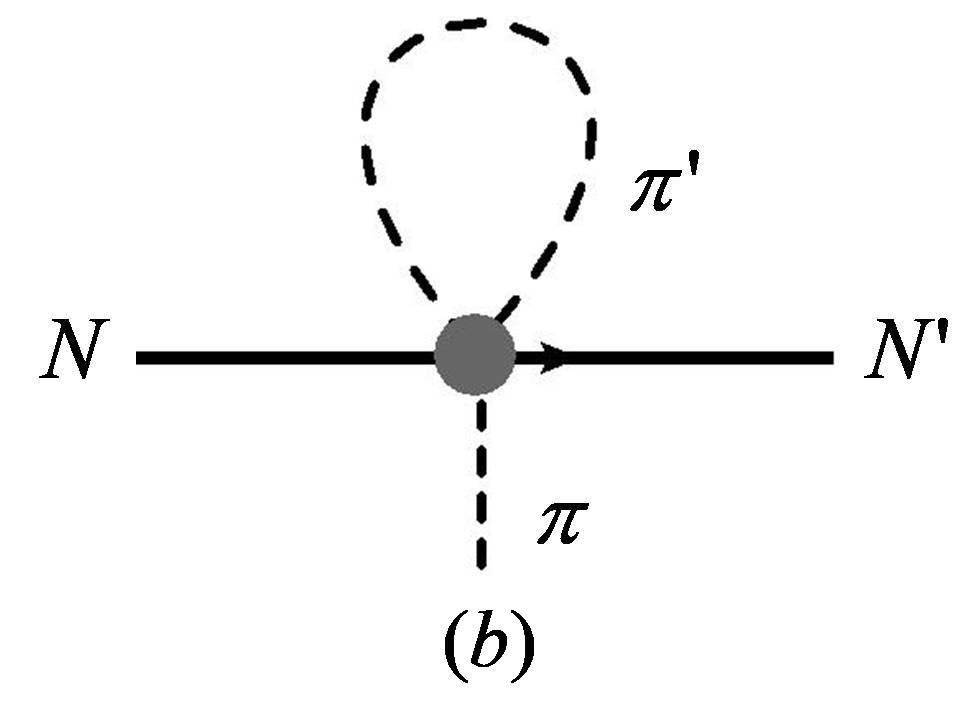}\hfill
     \includegraphics[width=0.32\linewidth]{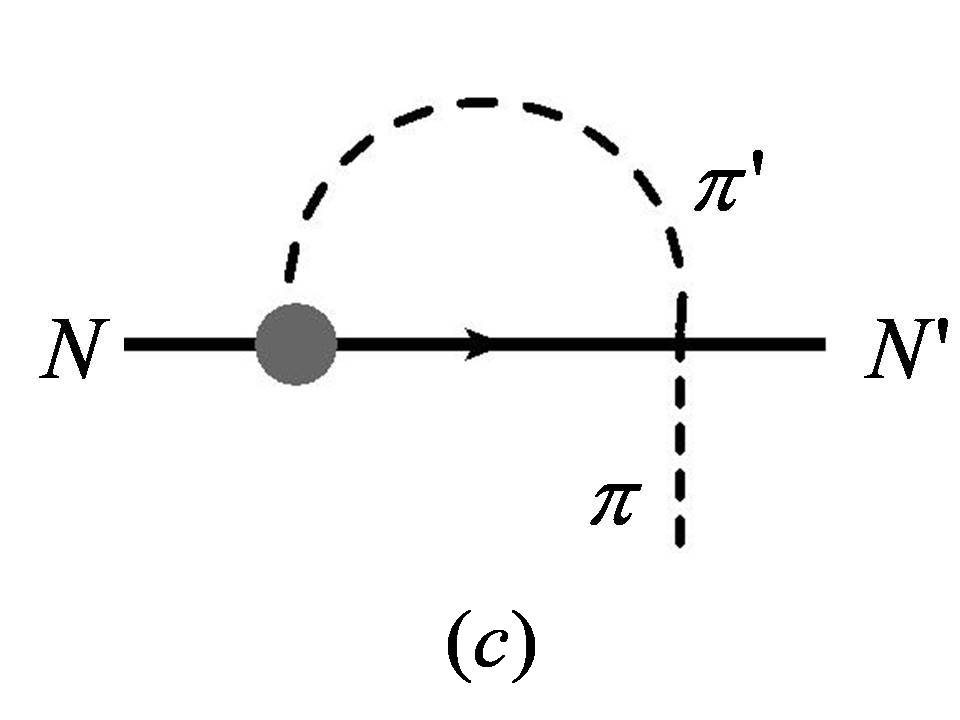}\\
     \includegraphics[width=0.32\linewidth]{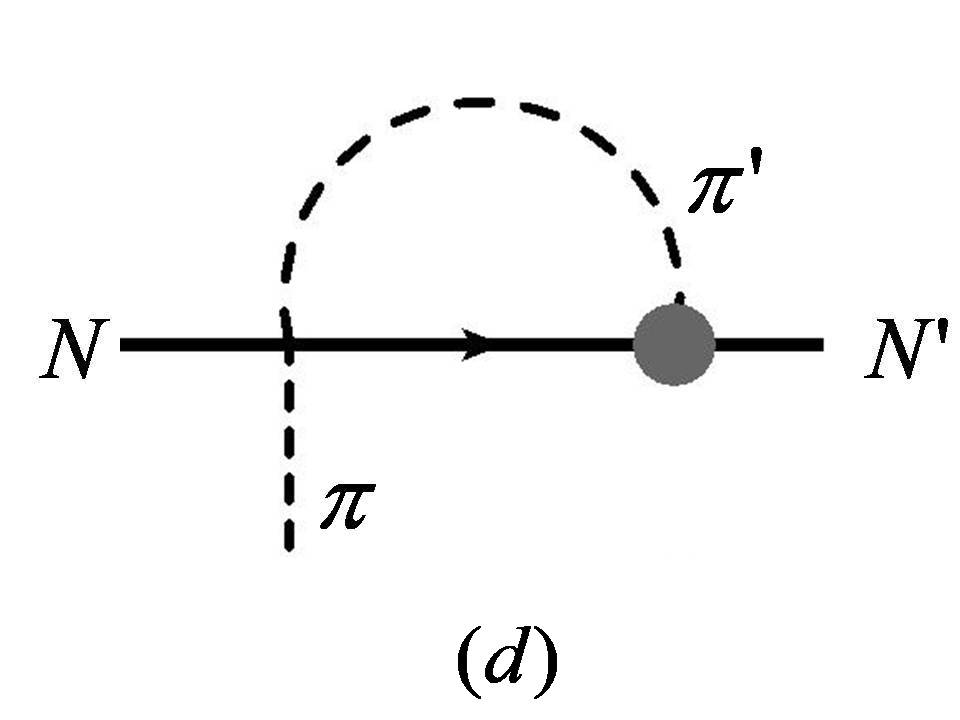}\hfill
     \includegraphics[width=0.32\linewidth]{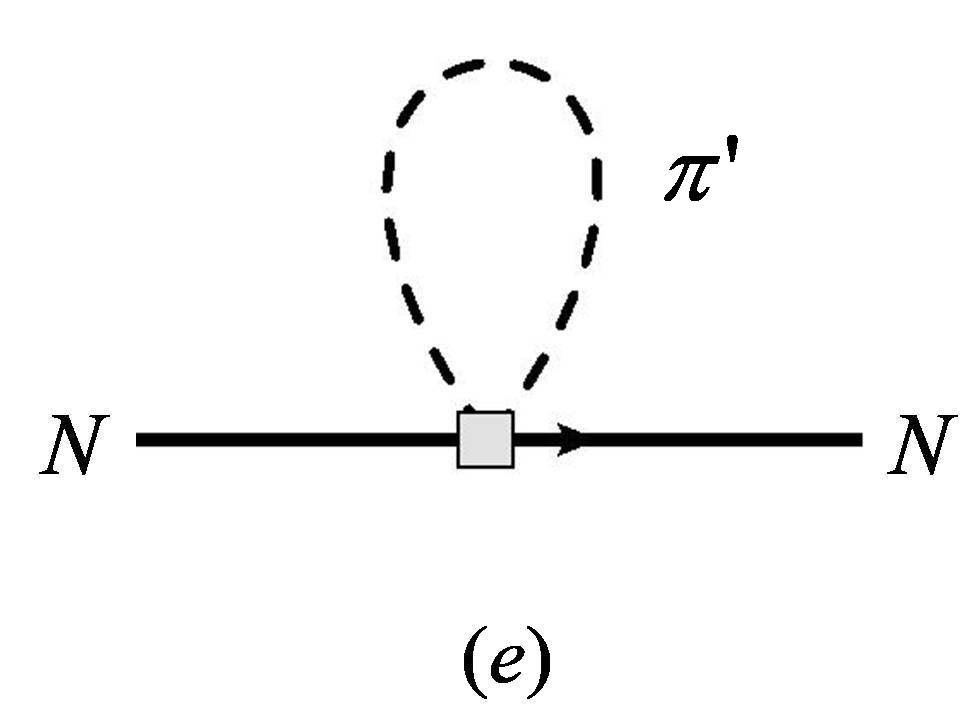}\hfill
     \includegraphics[width=0.32\linewidth]{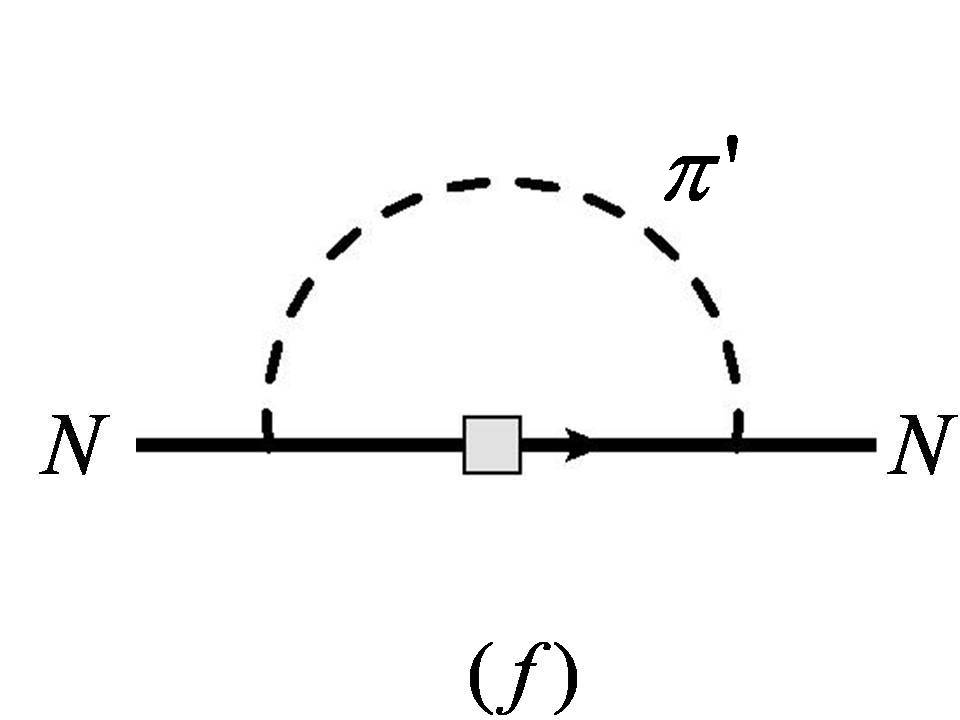}
	\par\end{centering}
	\caption{\label{fig:loop}One-loop diagrams that contribute to 
$h_\pi^1$((a)-(d)) and $(\delta m_N)_{4q}$ ((e)-(f)). The black dot and grey 
box denote the $\Delta I=1$ $NN\pi$ and $NN$ weak vertex insertion, 
respectively. Self-energy diagrams are not shown explicitly.}
\end{figure}

Equation~\eqref{eq:central} is the central result of our paper and it is derived from the LO ChPT. Next we consider {\color{black}the NLO effects} due to both one-loop diagrams and LECs to this tree-level relation. 
The relevant one-particle irreducible (1PI) loop diagrams are depicted in 
Fig.~\ref{fig:loop}. Meanwhile, one also needs to compute the renormalization 
of 
$F_\pi$ as well as the wavefunction renormalization of the pion field. They are 
given by
\begin{eqnarray}
\delta (F_\pi)=
-\frac{I_e}{F_\pi^2}+\frac{M_\pi^2}{F_\pi}l_4\,,~
\sqrt{Z_\pi}-1 = \frac{I_e}{3F_\pi^2}-\frac{M_\pi^2}{F_\pi^2}l_4\,,~
\end{eqnarray}
where $l_4$ is the well-known $O(p^4)$ LEC in the SU(2)-mesonic chiral 
Lagrangian~\cite{Gasser:1983yg}. The total 1-loop correction to 
the left- and right-hand-side of 
Eq.~\eqref{eq:central}, 
including both the 1PI and wavefunction renormalization contributions, 
reads\footnote{{\color{black}The one-loop corrections to $h_\pi^1$ was previously calculated in 
\cite{Zhu:2000fc}. }
% with the inclusion of explicit $\Delta$-resonances. Here we 
% do not include them because the main conclusion of this paper, namely whether 
% or not the matching relation is preserved at one loop, is not altered with or 
% without the $\Delta$s.  
}
\begin{eqnarray}
\delta\left(F_\pi 
h_\pi^1\right)_{\mathrm{loop}}\ali=\ali 
\left(\frac{g_A^2}{F_\pi^2}I_a-\frac{1}{F_\pi^2 }I_e+\delta Z_N\right)F_\pi 
h_\pi^1\,,\\
\delta\left((\delta 
m_N)_{4q}\right)_{\mathrm{loop}}\ali=\ali 
\left(\frac{g_A^2}{F_\pi^2}I_a-\frac{1}{ F_\pi^2}I_e+\delta Z_N\right)(\delta 
m_N)_{4q}\,,\nonumber\label{eq:loopcor}
\end{eqnarray}
where
\begin{eqnarray}
I_{a} =
- \frac{3M_{\pi}^{2}}{64\pi^{2}}\left(R_\pi+\frac{2}{3}\right), ~~
I_{e} =
\frac{M_{\pi}^{2}}{16\pi^{2}}R_\pi,
\end{eqnarray}
with $R_\pi=\frac{2}{d-4}+\gamma_{E}-\ln(4\pi)-1+\ln({M_
{\pi}^{2}}/{\mu^{2}})$ and $\mu$ the renormalization 
scale,
are the loop functions defined in Ref.~\cite{Seng:2016pfd}, and $Z_N=1+\delta 
Z_N$ is 
the nucleon wavefunction renormalization whose explicit form does not concern 
us. From Eq.~\eqref{eq:loopcor} one observes that the loop corrections to both 
sides of Eq.~\eqref{eq:central} simply result in a common multiplicative 
factor, so the matching relation is unaltered by one-loop corrections. The 
results above are obviously incomplete because one needs to introduce the NLO 
weak chiral Lagrangian simultaneously in order to absorb the ultraviolet 
divergences in the loop 
diagrams as well as to render the final expressions scale-independent. Such a
Lagrangian involves a single insertion of the quark mass matrix. There are only 
two independent terms at this order~\cite{Beane:2002ca},
\begin{equation}
\mathcal{L}^w_{\mathrm{tot,NLO}}=\tilde{c}_1\bar{N}\left\{\chi_+,X_R\right\}
N+\tilde{c}_2\mathrm{Tr}\left(\chi_+\right)\bar{N}X_RN.
\end{equation}
Their contributions are
\begin{eqnarray}
\delta(h_\pi^1)_{\mathrm{LEC}}\ali=\ali 
-\frac{8\sqrt{2}}{F_0}B_0\bar{m}(\tilde{c}_1+\tilde{c}_2)\,,\nonumber\\
\delta((\delta m_N)_{4q})_{\mathrm{LEC}}\ali=\ali 
16B_0\bar{m}(\tilde{c}_1+\tilde{c}_2),\label{eq:LwNLO}
\end{eqnarray} 
where $\bar{m}=(m_u+m_d)/2$. We find that the 
quantities $\delta(h_\pi^1)_{\mathrm{LEC}}$ and $\delta((\delta 
m_N)_{4q})_{\mathrm{LEC}}$ also satisfy the matching relation in 
Eq.~\eqref{eq:central}. Therefore, the soft-pion theorem relating
$h_\pi^1$ to
$(\delta m_N)_{4q}$ is protected against all corrections {\color{black}of NLO, including 
both 
the one-loop and LEC contributions and without assuming isospin symmetry}. Hence, we expect the accuracy of such a relation to be better than 
$(M_\pi/\Lambda_\chi)^2\sim1\%$ when the light quark 
masses take the physical values.

In the conventional approach, the $h_\pi^1$ coupling is extracted from the 
hadronic matrix element involving the nucleon-pion state. 
Such a hadronic matrix element can be calculated nonperturbatively using 
lattice QCD. However, it would cause three complexities: 

\begin{itemize}
\item The nucleon-pion state is a rescattering state. As a result, the hadronic 
matrix element calculated in the finite lattice box suffers from a power-law 
finite-volume effect. Only after the appropriate finite-volume
correction~\cite{Lellouch:2000pv} can
the hadronic matrix element in the finite volume be related to 
the physical one in the infinite volume. 

\item Because of the inequality of the energies of the on-shell nucleon and the
nucleon-pion state, the weak four-quark operator involves an energy insertion.
As the injected energy must exceed $E\ge M_\pi+m_n-m_p$, the LO
effect of the energy insertion scales as $\sqrt{m_q}$ and may dominate 
over the loop correction in ChPT. This LO
contamination can be removed by antisymmetric combination of the forward $(p\to 
n\pi)$  
and backward $(n\pi\to p)$ transitions~\cite{Beane:2002ca}, but the higher 
order terms $\sim m_q$  
still remain. Although the systematical effect associated with 
the non-zero energy insertion vanishes in the chiral limit, it still complicates 
the lattice 
calculation as non-zero quark masses are used in the simulation. 

\item Although a three-quark interpolating operator can be used to create 
the nucleon-pion state in the $S$-wave~\cite{Wasem:2011zz}, it is known from  
lattice QCD 
that the overlapping amplitude between a three-quark operator and a two-hadron 
state can be significantly suppressed~\cite{Dudek:2012xn}. To gain a better 
precision, it is desirable to use a nucleon-pion 
interpolating operator to build the correlation function. However, it would 
make the quark contractions more complicated and the calculation more expensive.
\end{itemize}

By relating the P-violating hadronic matrix element to the P-conserving one 
using Eq.~\eqref{eq:central}, one can reduce a  
nucleon-pion state to a single nucleon state. As a consequence all of the three 
complexities mentioned above disappear, and the calculation is much simplified.
Considering the fact that the only-existing lattice calculation performed 
at $M_\pi\simeq 389$~MeV yields a result with $\sim$50\% statistical 
uncertainty~\cite{Wasem:2011zz}, it is an important intermediate step to study 
the P-conserving matrix element as an alternative. Using the 
Feynman--Hellmann method proposed in Ref.~\cite{Bouchard:2016heu}, one can 
calculate $(\delta m_N)_{4q}$ using the correlation functions of a single time 
variable, which simplifies the procedure to remove the excited-state 
contamination.

Finally, the soft-pion theorem shown here also brings benefits to other 
diagram-based analysis of $h_\pi^1$ such as Dyson-Schwinger Equation and 
partially-quenched ChPT. The former, for example, computes hadronic matrix 
elements by evaluating ``loop diagrams" of quark and gluons with insertions of 
fully-dressed vertices and propagators obtained by solving integral equations. 
The disappearance of pion in the external state greatly reduces the number of
diagrams, in particular those involving contractions between the quarks in 
the nucleon and the pion. The latter is able to isolate 
diagrams with specific contractions in a given hadronic matrix element through 
calculations of tree and loop diagrams in ChPT with an extended flavor 
sector~\cite{DellaMorte:2010aq,Juttner:2011ur,Acharya:2017zje}. Based on our 
theorem, the objects of interest are translated to baryon mass parameters, so 
the number of loop diagrams is much smaller (which can be seen from Fig. 
\ref{fig:loop}), making the calculation more tractable.

In summary, we demonstrate that the $h_\pi^1$, induced by $\Delta I=1$  
parity-odd four-quark operators resulting from the exchange of the $Z$-boson as 
well 
as QCD running, can be related to the neutron-proton mass splitting $(\delta 
m_N)_{4q}$ induced by a corresponding set of $\Delta I=1$, parity-even 
four-quark operators. The matching relation is established as a 
soft-pion theorem $F_\pi h_\pi^1=-(\delta m_N)_{4q}/\sqrt{2}$, which is 
protected against any correction {\color{black} of NLO} in the chiral expansion. 
Therefore, instead of $h_\pi^1$ one may study $(\delta m_N)_{4q}$ which is a 
much simpler hadronic matrix element due to the disappearance of pion from the 
external state. Such matching brings about benefits to both lattice and other 
nonperturbative QCD calculations of $h_\pi^1$. We hope that our finding could 
serve as a new starting point for the next round of theoretical investigations 
which could be directly contrasted 
to the upcoming experimental results, and thus shed new lights on the many 
unresolved puzzles in hadronic weak interactions. 

\bigskip

We thank Emanuele Mereghetti and Michael Ramsey-Musolf for many useful 
discussions. We are particularly indebted to Jordy de Vries who provided many 
useful suggestions as well as contributed to the refinement of the introduction. 
The work of CYS is supported in part by the National Natural Science Foundation 
of China (NSFC)
under Grant Nos.11575110, 11655002, 11735010, Natural Science Foundation of 
Shanghai under Grant No.~15DZ2272100 and No.~15ZR1423100,  by Shanghai Key 
Laboratory for Particle Physics and Cosmology,  and by Key Laboratory
for Particle Physics, Astrophysics and Cosmology, Ministry of Education; he 
also appreciates the support through the Recruitment Program of Foreign Young 
Talents from the State Administration of Foreign Expert Affairs, China.
The work of XF is supported in part by NSFC under Grant No. 11775002.
The work of FKG is supported in part by DFG and NSFC through funds provided to 
the Sino-German CRC 110 ``Symmetries and the Emergence of Structure in QCD" 
(NSFC Grant No.~11621131001, DFG Grant No.~TRR110), by NSFC under Grant 
No.~11747601, by 
the CAS Key Research Program of Frontier Sciences (Grant No.~QYZDB-SSW-SYS013), and by the CAS Center for Excellence in Particle Physics (CCEPP). 
XF and FKG also thank support provided through the Thousand Talents Plan for 
Young Professionals.

\bibliography{hpi_ref}

\end{document}